\documentclass[twocolumn,11pt]{article}
\usepackage[a4paper,margin=0.75in]{geometry}
\usepackage{amsmath,amssymb}
\usepackage{graphicx}
\usepackage{booktabs}
\usepackage{multirow}
\usepackage{makecell}
\usepackage{array}
\usepackage{caption}
\usepackage{xspace}
\usepackage[hidelinks]{hyperref}
\usepackage{authblk}
\usepackage{indentfirst}
\graphicspath{{figures/}}

\newcommand{\us}{\mu\mathrm{s}}

\newcommand{\MeVc}{\mathrm{MeV}/c}

\title{Neural-Network-Assisted Binary Template Construction for Matrix-Based Pattern Matching in the STCF MDC}

\author[1]{Zhaoli Guo}
\author[1]{Qingyu Li}
\author[1]{Baolin Zhang}
\author[1]{Jiarui Zhao}
\author[1,2]{Aonan Zhu}
\author[1,2]{Liang Peng$^{*}$}
\author[1,2]{Huilin Li$^{**}$}

\affil[1]{School of Physics and Electronic Science, Hunan University of Science and Technology, Xiangtan 411201, China}
\affil[2]{Department of Electronic Science and Technology, Hunan University of Science and Technology, Xiangtan 411201, China}
\affil[ ]{$^{*}$Corresponding author: pengliang@hnust.edu.cn}
\affil[ ]{$^{**}$Corresponding author: lihuilin@hnust.edu.cn}

\date{\today}

\begin{document}
	
	% ============================================================
	% Title + abstract in one-column width
	% ============================================================
	\twocolumn[
	\maketitle

	\begin{abstract}

		The Super Tau-Charm Facility, operating at high luminosity, will produce high event rates and high data throughput, imposing stringent requirements on fast track finding and data reduction and compression algorithms in the High-Level Trigger. Local track segment finding in the Main Drift Chamber underpins subsequent segment combination and full track reconstruction, yet high background rates and limited detection efficiency can significantly increase the risk of false triggers and signal loss in pattern matching algorithms. This paper presents a neural-network-assisted framework for constructing binary template libraries used in matrix-based pattern matching for MDC local track segment finding. The framework formulates template construction as a differentiable multi-objective optimization problem, employing a neural network to jointly learn template parameters under multiple constraints. After training, only binary template pairs are exported and deployed into the existing bitwise pattern matching routine, requiring no neural network inference at runtime and thus preserving the deterministic, fast, and parallelizable nature of the online algorithm. Experimental results based on simulation samples demonstrate that, under limited detection efficiency, the resulting template library maintains relatively high signal retention across different transverse momentum ranges and background levels, and can be flexibly tailored to adjust the coverage range according to practical requirements. The proposed approach decouples the physics performance from the computational speed by combining the improved physics performance brought by offline neural-network-based optimization with the determinism and high speed of a conventional online algorithm, suggesting a new research direction for artificial-intelligence-enhanced online data processing in high-luminosity particle collider experiments.
	
%		\noindent\textbf{Keywords:} Super Tau-Charm Facility; main drift chamber; high-level trigger; pattern matching; binary template; neural network
		\end{abstract}

	]
	
	% ============================================================
	\section{Introduction}
	\label{sec:introduction}
	
	The Super Tau-Charm Facility (STCF) is a next-generation large-scale electron--positron collider facility currently planned in China~\cite{STCFdesign}. Designed for the tau-charm energy region, the STCF is expected to operate at a center-of-mass energy of
	\(\sqrt{s}=2\)--\(7~\mathrm{GeV}\), with a peak luminosity exceeding \(0.5\times10^{35}~\mathrm{cm}^{-2}\mathrm{s}^{-1}\). This energy region covers important threshold domains for light hadrons, charmed hadrons, and tau leptons, providing a crucial window for studying non-perturbative effects in quantum chromodynamics, performing precision tests of the Standard Model, and searching for new physics beyond the Standard Model.

	When the STCF operates in its highest designed luminosity regime, the raw data throughput is expected to exceed \(30~\mathrm{GB/s}\). Directly writing all raw data to offline storage would not only impose substantial pressure on data transmission and storage, but also increase the computational burden of subsequent offline reconstruction and physics analysis. Therefore, effective event selection, reconstruction, and compression at the online stage are among the key requirements for the successful operation of the STCF experiment.
	
	In modern high-energy collider experiments, a trigger and data acquisition (TDAQ) system is commonly employed~\cite{ATLASTDAQ,BESIII,BelleIIDAQ}. Such a system progressively reduces the enormous raw data volume to an acceptable level through multi-level triggering and hierarchical data processing. Similar to experiments such as ATLAS and CMS~\cite{ATLASHLT,ATLASHLT2,CMSHLT}, the STCF also plans to adopt a multi-level trigger system. In the STCF, the TDAQ system mainly consists of the Level-1 trigger (L1), the data acquisition system (DAQ), and the high-level trigger (HLT)~\cite{STCFdesign}. The digitized signals produced by the detector are sent simultaneously to the L1 and the DAQ. The L1 performs rapid decisions on collision signals using low-latency hardware algorithms~\cite{STCFL1trigger,STCFL1trigger2}. When the L1 identifies an event as potentially physics-relevant, it sends an accept command to the DAQ. The DAQ receives another copy of the detector signals and, together with the L1 decision, assembles the complete event and transmits it to the HLT through the network system. As a software trigger system running on a high-performance online computing cluster, the HLT can exploit more complete detector information and more sophisticated software algorithms to further select, reconstruct, and compress events. After HLT processing, the selected data are finally recorded into the offline database.

	Within the overall trigger chain, the L1 and the HLT are responsible for tasks at different levels. The L1 is designed for low-latency hardware-based decisions and must directly reject a large number of irrelevant events under stringent timing constraints. It is therefore suitable for implementation on hardware platforms such as field-programmable gate arrays (FPGAs) and application-specific integrated circuits (ASICs). In contrast, the HLT needs to inspect events in greater detail and, while effectively suppressing background hits, preserve and reconstruct true particle signals as completely as possible. To accomplish more complex reconstruction and selection tasks, the HLT necessarily relies on more sophisticated software algorithms. These algorithms can be data-dependent and may involve conditional branches, dynamic data structures, and irregular memory accesses. Therefore, the current HLT design plans to deploy such algorithms on general-purpose processors.
	
	The charged-track information from the main drift chamber (MDC) is one of the fundamental inputs for event reconstruction and selection in the HLT. Local segment finding not only provides seeds for subsequent segment combination, but also gives coarse information on the position, curvature direction, and momentum range of candidate segments. Such local information forms the basis for subsequent full track reconstruction. The reconstructed charged tracks can then be extrapolated inward and outward to support vertex constraints and the association of responses from subdetectors such as particle identification systems and the electromagnetic calorimeter. Under high-luminosity operating conditions, MDC hits are simultaneously affected by background hits and finite detection efficiency, which further increases the difficulty of local segment identification.
	
	Pattern matching refers to a class of algorithms that determine whether a matching condition is satisfied, or evaluate the degree of similarity, by comparing the features and spatial relationships between an object and predefined patterns~\cite{PAT1,PAT2,PAT3,PAT4}. In drift-chamber track reconstruction, pattern matching methods are commonly used for fast local segment finding. In a pattern matching algorithm, the online processing speed is mainly determined by the data layout of the program, the parallelization strategy, and the number of templates, whereas the segment identification performance is mainly determined by the coverage, robustness, and background-suppression capability of the template library. The quality of the template library directly determines the efficiency, purity, and output data volume of local segment finding. Therefore, constructing a compact, robust, and high-quality template library is essential for improving the performance of the matrix-based pattern matching algorithm.

	This paper mainly investigates how to use a neural network as an offline differentiable optimization tool to construct a high-quality binary template library for the matrix-based pattern matching algorithm. We first briefly review the matrix-based pattern matching algorithm. This algorithm maps each MDC superlayer into a two-dimensional matrix and finally converts it into one-dimensional data, and then performs fast segment finding and associated hit recovery through intersection and union operations implemented by the arithmetic logic unit (ALU). However, the template construction method used in this algorithm and the strict bitwise matching procedure are susceptible to the combined effects of finite detector efficiency and background hits. In addition, the template grouping problem for different local-segment morphologies is complex. Therefore, this paper introduces a neural network~\cite{HubaraBNN} as a differentiable optimization tool to learn a binary template library suitable for matrix-based pattern matching. The resulting template library enables the matrix-based pattern matching algorithm to maintain a high signal retention rate, a low output fraction, and a high signal retention ratio and a low output fraction under high-background and reduced-efficiency conditions. On the one hand, the online algorithm can remain fixed and efficient, which is particularly important for trigger and online reconstruction applications where latency, stability, and reproducibility are critical. On the other hand, the template library can be independently optimized in an offline stage, where artificial-intelligence-based methods can be introduced without affecting the online processing latency. The method proposed in this paper introduces artificial intelligence techniques into a classical trigger algorithm, decoupling the execution efficiency of the matching algorithm from the optimization complexity of the template construction problem. As a result, a fast, rule-based online local segment finding algorithm is obtained. This provides a useful perspective for combining machine-learning-based optimization with traditional online algorithms in high-luminosity particle physics experiments.

	% ============================================================
	\section{STCF MDC and the Matrix-Based Pattern Matching Algorithm}
	\label{sec:stcf_mdc}
	
	The main drift chamber (MDC) is an essential component of the tracking system of the STCF detector. It is designed as a cylindrical multi-wire drift chamber. When charged particles traverse the working gas, they ionize gas molecules, and the generated ionization electrons drift toward the sense wires under the electric field, producing readout electrical signals. Using the hit information from multiple detector layers, the spatial trajectories of charged particles can be reconstructed, from which physical quantities such as momentum can be further obtained.
	
	The MDC consists of 48 layers of drift cells in total, with every six layers forming one superlayer. The number of cells is identical for all layers within the same superlayer. The entire tracking system is placed in a uniform magnetic field of \(1~\mathrm{T}\) along the \(z\)-axis, with a polar-angle coverage of \(20^\circ\)--\(160^\circ\). The MDC contains both field wires and sense wires. The sense wires collect signals induced by the ionization produced by charged particles. They are mainly divided into axial wires and stereo wires, where the axial wires are parallel to the beam direction and the stereo wires are arranged at a certain angle with respect to the beam direction. Information such as timing and charge collected from these signals provides the basic input for subsequent track reconstruction and particle identification.
	
	Under the high-luminosity operating conditions of the STCF, the raw hit data from the MDC are affected by complex background processes. These backgrounds can generally be classified into beam-related backgrounds and luminosity-related backgrounds. The beam-related backgrounds mainly include the Touschek effect and beam--gas interactions, including Coulomb scattering and bremsstrahlung, whereas the luminosity-related backgrounds mainly include radiative Bhabha scattering and two-photon processes~\cite{OSCAR3,STCFBackground}. Since the MDC responds to charged particles passing through its sensitive volume, both background particles and true charged particles produced by the target physics processes can generate hits on the sense wires.

	In a gaseous drift chamber, the formation of valid hits is jointly affected by several processes, including particle ionization, gas amplification, and electronic readout. Owing to statistical fluctuations in primary ionization and signal amplitude, some signals may fall below the electronic threshold. In addition, if the drift time lies outside the valid readout time window, the signal may not be recorded as a valid hit. Dead channels, local inefficiency regions, and geometric boundary effects can also lead to a local decrease in hit efficiency~\cite{BESIIIAging}. Under high-luminosity operating conditions, beam-related backgrounds increase the channel occupancy, thereby causing signal pileup and electronic dead time, which further result in the loss of valid hits.

	Fast track finding is a key task in the trigger and reconstruction systems of high-luminosity collider experiments. In traditional approaches, algorithms such as template matching, associative memories, and cellular automata have been widely used for fast track recognition, track-seed finding, and local segment identification~\cite{STCFL1trigger,JiaSegment,KiselCBM}. These methods usually have regular computational structures and are suitable for online environments and parallel hardware implementation. However, their performance often depends on the design and tuning of pattern banks, template libraries, or local-rule parameters. In recent years, machine learning methods have also been extensively investigated for track finding and hit association, including approaches based on multilayer perceptrons, convolutional neural networks, and graph neural networks~\cite{JuGNN,BelleIIGNN}. These methods provide flexible optimization tools for pattern recognition in complex detector environments, but their deployment overhead, latency, and robustness must be carefully considered according to the specific trigger or reconstruction scenario.

	This study adopts a strategy different from direct online neural-network inference. The neural network is used only in the offline stage to optimize the binary template library. After training, the exported templates are deployed in the matrix-based pattern matching algorithm. The online stage still uses deterministic binary template matching operations without executing neural-network inference. In this way, the optimization of template parameters is decoupled from the execution of online matching. The offline training stage is not directly constrained by online trigger latency and can therefore be used to sufficiently optimize the template library, while the online stage preserves the regular, deterministic, and low-latency execution form of the matrix-based pattern matching algorithm.
	
\begin{figure}[tbp]
		\centering
		\includegraphics[width=0.98\columnwidth]{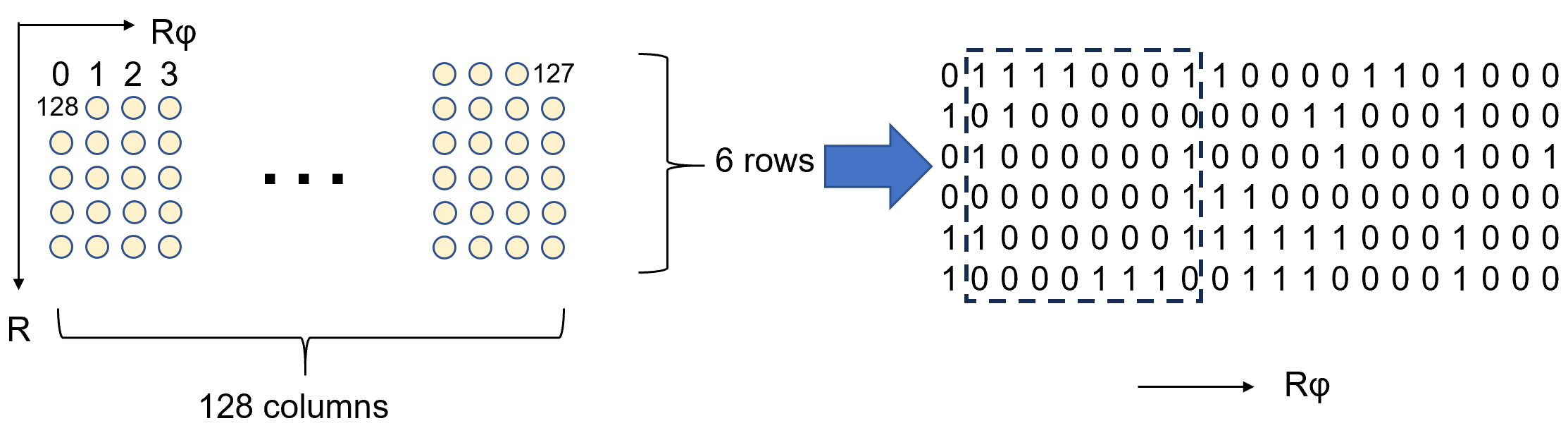}
		\caption{Each superlayer of the MDC is mapped to a two-dimensional Boolean matrix, where the sliding window indicated by the dashed box moves along the direction of increasing column index.}
		\label{fig:MDCtoBool}
	\end{figure}

	Each MDC superlayer is represented as a two-dimensional Boolean matrix with a fixed number of rows, where each matrix element indicates whether the corresponding sense wire is hit in a given event. By sliding a fixed-size window along the column direction, local segment identification can be transformed into a structured pattern matching problem. Figure~\ref{fig:MDCtoBool} illustrates this process, where the sliding window traverses these matrices for pattern matching.
	
%	To reduce the template scale, preserve hit-recovery capability, and suppress false triggers caused by dense background signals, we design a trigger--recovery separation strategy. After grouping templates corresponding to similar tracks, two types of template objects are constructed: the trigger template \(\Gamma\) and the recovery template \(\Delta\). Each template pair consists of one trigger template and one recovery template, which are used for candidate local segment triggering and associated hit retention, respectively. In the trigger stage, a skeleton-like core pattern is used to determine whether the local observation window matches a candidate configuration. In the recovery stage, an envelope-like region is used to retain information associated with the track. This strategy alleviates the scale pressure caused by the ``one track, one template'' scheme without significantly increasing the complexity of online operations.
	
\begin{figure}[t]
		\centering
		\includegraphics[width=\columnwidth]{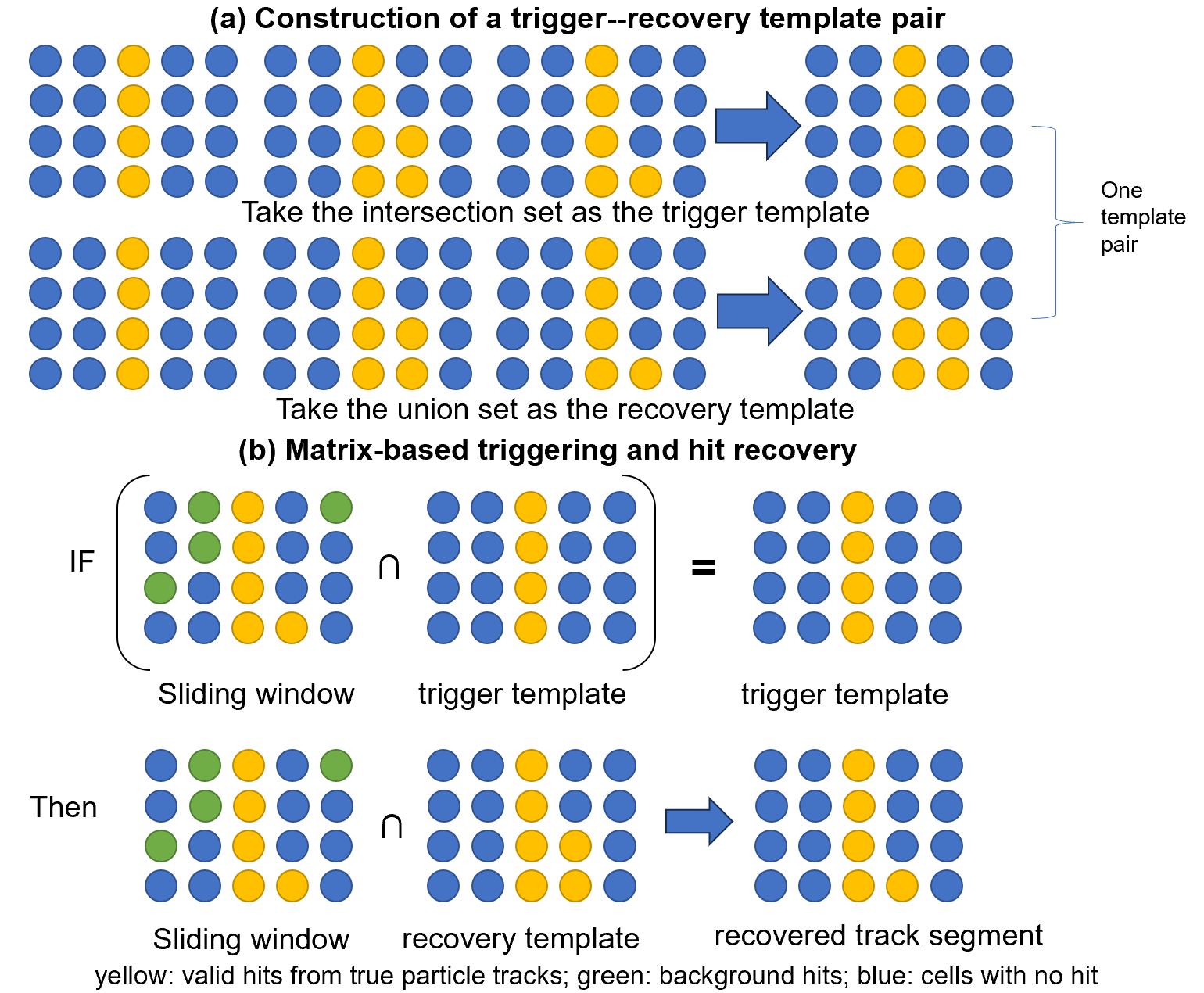}
		\caption{
			Schematic illustration of the trigger--recovery separation strategy.
			(a) A trigger--recovery template pair is constructed from a group of templates corresponding to similar tracks. The trigger template is obtained from the intersection set as a skeleton-like core pattern, while the recovery template is obtained from the union set as an envelope-like region for hit retention.
			(b) For a sliding window, the template pair is triggered if its intersection with the trigger template is identical to the trigger template. The recovered track segment is obtained by intersecting the sliding window with the recovery template.
			Yellow, green, and blue circles denote valid hits from true particle tracks, background hits, and cells with no hit, respectively.
		}
		\label{fig:trigger_recovery}
	\end{figure}

	To reduce the template scale, preserve hit-recovery capability, and suppress false triggers caused by dense background signals, we design a trigger--recovery separation strategy, as illustrated in Fig.~\ref{fig:trigger_recovery}. After grouping templates corresponding to similar tracks, two types of template objects are constructed: the trigger template \(\Gamma\) and the recovery template \(\Delta\). Each template pair consists of one trigger template and one recovery template, which are used for candidate local segment triggering and associated hit retention, respectively. In the trigger stage, a skeleton-like core pattern is used to determine whether the local observation window matches a candidate configuration. In the recovery stage, an envelope-like region is used to retain information associated with the track. This strategy alleviates the scale pressure caused by the ``one track, one template'' scheme without significantly increasing the complexity of online operations.

	In this paper, the fixed-size sliding window is referred to as the local observation window. The trigger template \(\Gamma\) denotes the binary core hit pattern used to determine whether a local observation window contains a candidate local segment, whereas the recovery template \(\Delta\) denotes the binary envelope region used to retain relevant hits within the window after the template is triggered. The rigorous mathematical definitions of these two templates will be given in Section~\ref{sec:template_framework}.
	
	As described in the Introduction, the quality of the template library still determines the upper bound of the final physics performance. If the trigger template \(\Gamma\) is too sparse, the trigger condition may become overly loose, leading to an increased false-trigger rate from background hits. If different \(\Gamma\) templates overlap too strongly, multiple templates may be triggered simultaneously in a single event, thereby reducing selectivity. If the recovery template \(\Delta\) is too wide, the recovery envelope will introduce more background hits and worsen false coverage. In addition, under reduced detector-efficiency conditions, the templates must be sufficiently robust to avoid missing true tracks.
	
	Before detector efficiency is taken into account, the interference mainly comes from background hits. In this case, it is generally sufficient to select suitable tracks for grouping, generate the corresponding \(\Gamma\)--\(\Delta\) template pairs, and evaluate them on a large test set to identify the best combination. However, under realistic operating conditions, finite detector efficiency can lead to incomplete or fragmented local segments. Since the trigger condition of a \(\Gamma\) template relies on strict bitwise operations, the loss of even a small fraction of hits may cause the matching to fail. Therefore, detector inefficiency poses a major challenge to template construction. Random signal loss caused by detector inefficiency may occur at any position. Even for relatively simple high-momentum local segments, multiple templates may be required to satisfy the matching condition. In contrast, overly sparse templates may be triggered by dense background hits in localized regions of a superlayer, leading to false triggers in the algorithm.
	
	Based on this observation, we attempted to avoid such failures by substantially expanding the template library with incomplete \(\Gamma\) templates in which a small number of hits are missing. However, this approach leads to a large increase in the number of required templates. In addition, containment relationships may arise among different \(\Gamma\) templates, resulting in substantial redundancy in the template library. Therefore, under the combined effects of background hits and finite detector efficiency, the template grouping problem can no longer be effectively solved by manual selection and large-scale testing. Instead, it becomes a high-dimensional, multi-objective combinatorial optimization problem.
	
	To address this issue, we introduce a neural network as a template construction tool. Its differentiable optimization capability is used to learn template parameters. After training, binary templates are exported and deployed in the original matrix-based pattern matching framework. The use of a neural network is motivated by the following considerations. First, the direct optimization of binary templates is essentially a discrete combinatorial optimization problem. A neural network provides a continuous-relaxation approach, in which template responses can first be learned in a continuous parameter space and then binary \(\Gamma\) templates are exported through a discretization procedure. Second, the optimization of the template library must jointly consider multiple metrics, including local segment identification, the number of templates, and inter-template redundancy. By constructing a joint loss function, the neural network can integrate these objectives into a unified differentiable optimization framework, without relying on manual tuning of individual terms. Finally, the computational overhead of the neural network appears only in the training stage. During online deployment, only the exported binary templates are used to execute the original matrix-based pattern matching algorithm, thereby preserving the high-speed and deterministic advantages of matrix-based matching.
	
	% ============================================================
	\section{Neural-Network-Assisted Binary Template Construction Framework}
	\label{sec:template_framework}
	
	\subsection{Task Formulation and Template Semantics}
	\label{subsec:task_formulation}
	
	This section focuses on the generation of a binary template library, where each template has the same size as the fixed sliding window. We refer to this fixed sliding window as the local observation window. Each local observation window is represented as a Boolean matrix
	\begin{equation}
	M\in\{0,1\}^{H\times W},
	\end{equation}
	where the current implementation uses \(H=6\) and \(W=8\). A matrix element equal to 1 indicates that a valid hit exists in the corresponding cell, whereas an element equal to 0 indicates that no hit is present.
	
	To separately describe true tracks, detector-efficiency losses, and random backgrounds during training, three types of input matrices are defined in this work. \(M_{\rm clean}\) denotes an ideal local segment without background contamination or missing hits. \(M_{\rm miss}\) denotes an incomplete local segment obtained by randomly removing part of the true hits from \(M_{\rm clean}\), and is used to simulate missing hits caused by finite detector efficiency. \(M_{\rm noise}\) denotes a purely random noise background generated by a Bernoulli process, and is used to constrain the response of templates to background hits. In the current training implementation, detector-efficiency losses and random backgrounds are treated as two different input cases and enter different loss terms during training, rather than being explicitly combined into the same noisy observation window. It should be noted that these matrices do not represent three independent types of physical events. Instead, they are used during training to simulate different non-ideal conditions. The final exported binary templates are evaluated uniformly in the original hard-rule matching chain.
	
	For the \(k\)-th template, this work adopts a paired-template representation, consisting of a trigger template \(\Gamma_k\) and a recovery template \(\Delta_k\), where
	\begin{equation}
	\Gamma_k\in\{0,1\}^{H\times W},\qquad
	\Delta_k\in\{0,1\}^{H\times W}.
	\end{equation}
	The trigger template \(\Gamma_k\) describes the key hit structure required for triggering, and is expected to be supported as much as possible by incomplete local segments. The recovery template \(\Delta_k\) describes the recovery region allowed by this template, and is used to retain hits associated with the target track. In set notation, their ideal physical semantics can be written as
	\begin{equation}
	\Gamma_k\subseteq M_{\rm miss},
	\qquad
	M_{\rm clean}\subseteq \Delta_k .
	\end{equation}
	These inclusion relations are not required to be strictly satisfied during training, but are instead approximated through the loss function. After export, the binary templates are used in the hard-rule matching chain according to these template semantics. The former relation indicates that the active cells of the trigger template should still be supported by incomplete local segments as much as possible, whereas the latter indicates that the recovery template should cover the hits of the true track as completely as possible.
	
	In practical matrix-based matching, once a template is triggered, the final retained hits are those selected from the observation window and located inside the region defined by \(\Delta_k\):
	\begin{equation}
	R_k = M_{\rm obs}\cap \Delta_k ,
	\end{equation}
	where \(M_{\rm obs}\) denotes the observation window formed in realistic conditions by the superposition of an incomplete local segment affected by detector efficiency and random background hits. During training, however, detector-efficiency losses and pure-noise responses enter different loss terms through \(M_{\rm miss}\) and \(M_{\rm noise}\), respectively, rather than through their superposed observation window.
	
	\subsection{Network Architecture and Template Parameterization}
	\label{subsec:network_architecture}
	
	For each candidate template, the model optimizes a continuous parameter matrix of size \(6\times 8\). If there are \(K\) templates in total, the learnable parameters can be represented as
	\begin{equation}
	W\in\mathbb{R}^{K\times 1\times 6\times 8}.
	\end{equation}
	The continuous parameters are first mapped into probabilities through the sigmoid function,
	\begin{equation}
	P_{\Gamma}=\sigma(W),
	\end{equation}
	and then binarized with a threshold of 0.5 in the forward pass to obtain the binary core template \(\Gamma\). Since the binarization operation itself is non-differentiable, a straight-through estimator~\cite{YinSTE} is used during training to approximately pass gradients, enabling the model to optimize the templates by gradient descent while maintaining the binary template form.

\begin{figure*}[t]
		\centering
		\includegraphics[width=0.95\textwidth]{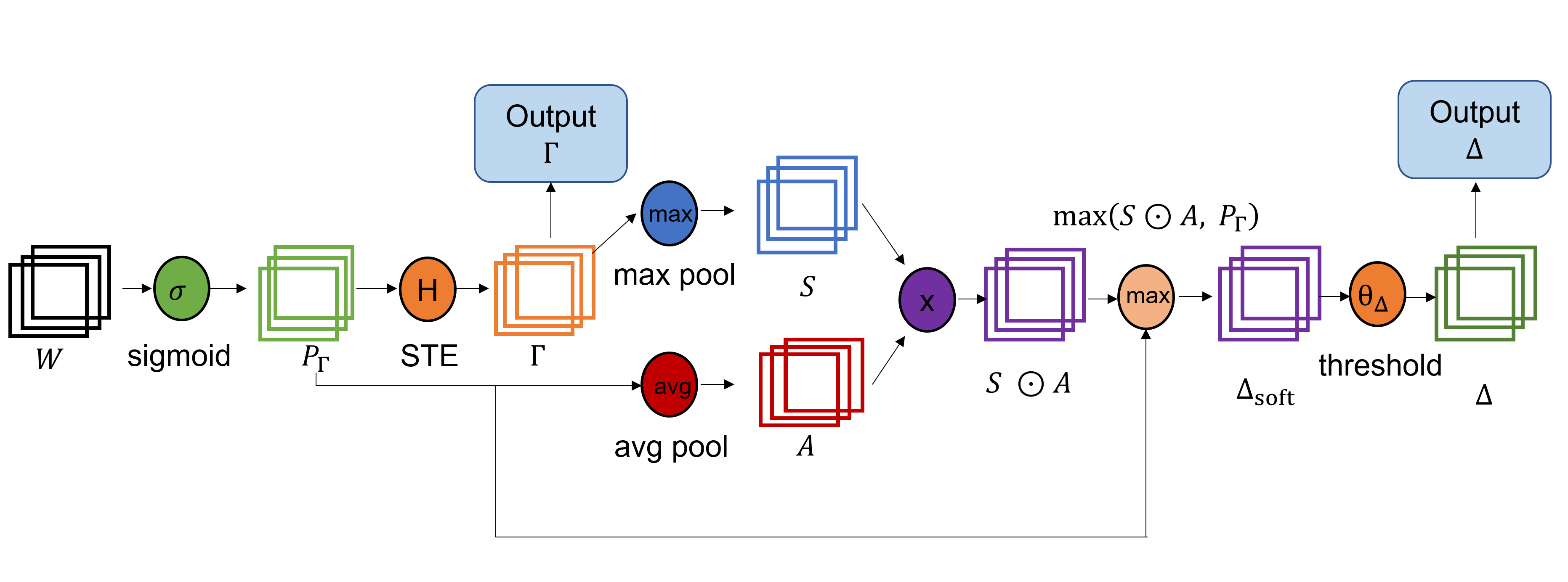}
		\caption{Architecture of the binary template network. The learnable weights $W$ are passed through a sigmoid function to obtain $P_{\Gamma}$, and binarized using a straight-through estimator to produce the binary trigger template $\Gamma$. The soft recovery template $\Delta_{\rm soft}$ is constructed by combining the dilated support region $S$ with the local average response $A$, and is then thresholded to obtain the binary recovery template $\Delta$.
		}
		\label{fig:binary_template_network}
	\end{figure*}

	The model architecture is shown in Fig.~\ref{fig:binary_template_network}. It should be noted that this work does not learn two independent sets of parameters. Instead, only \(\Gamma\) is treated as the learnable object, and \(\Delta\) is derived from \(\Gamma\). Specifically, a \(3\times 3\) local max-pooling operation is first applied to the binarized \(\Gamma\) to obtain the geometric support region of the template. Meanwhile, a \(3\times 3\) local average-pooling operation is applied to the continuous probability map \(P_{\Gamma}\) to obtain a smoothed local response. The two maps are then multiplied element-wise to form a soft \(\Delta\) template. Subsequently, the model takes the element-wise maximum between this soft template and \(P_{\Gamma}\), so that \(\Delta_{\rm soft}\) numerically preserves at least the information of the core trigger probability map:
	\begin{equation}
	\Delta_{\rm soft}
	=
	\max\left(S\odot A, P_{\Gamma}\right),
	\end{equation}
	where \(S\) denotes the support region obtained by dilating the binary \(\Gamma\), \(A\) denotes the response map obtained by local averaging of \(P_{\Gamma}\), and \(\odot\) denotes element-wise multiplication. This treatment helps preserve the core trigger region of \(\Gamma\) during final threshold-based export. However, it does not mean that \(\Delta_{\rm soft}\) is strictly constrained during training to be a hard template containing the binary \(\Gamma\). After training, \(\Delta_{\rm soft}\) is discretized into a binary recovery template \(\Delta\) using a fixed threshold. The resulting \(\Gamma\) and \(\Delta\) templates are fully compatible with the original matrix-based pattern matching algorithm.
	
	% Fig. 1 is the only double-column figure
	
	This construction makes the main response of \(\Delta\) jointly determined by the neighborhood support of \(\Gamma\) and its continuous probability response, and therefore tends to form a local recovery envelope around the trigger skeleton. Since the implementation further preserves the information of the core trigger probability map through \(\max(S\odot A,P_{\Gamma})\), this procedure mainly serves as a differentiable approximation to local expansion, rather than a strict hard constraint on the spatial range of \(\Delta\).
	
	After training, only the binary \(\Gamma\) and \(\Delta\) templates are exported and used as the new template library in the original matrix-based pattern matching workflow. Therefore, the computational overhead introduced by the neural network exists only in the offline training stage, while the online stage still preserves the deterministic and high-speed characteristics of the matrix-based pattern matching algorithm. The optimization objectives for the template parameters are described in the next section.
	
	% ============================================================
	\subsection{Optimization Objectives}
	\label{subsec:optimization_objectives}
	
	To jointly account for track retention, background suppression, and template redundancy during template construction, this work adopts a multi-objective joint loss function. The loss function directly reflects the physical requirements of the \(\Gamma/\Delta\) templates in matrix-based pattern matching. For each batch, true hits are randomly dropped with a given probability to obtain \(M_{\rm miss}\), which is used to simulate missing hits caused by finite detector efficiency. Meanwhile, a Bernoulli random-noise matrix \(M_{\rm noise}\) is generated to constrain the response of templates to random background hits. In the training loss, these two inputs are used separately to impose trigger-robustness and noise-rejection constraints, and are not explicitly superposed into the same training input window.
	
	For sample \(i\) and template \(k\), the matching cost is written as
\begin{equation}
	\begin{split}
		C_{ik}
		={} & w_{\rm sub}L^{ik}_{\rm sub}
		+ w_{\rm cov}L^{ik}_{\rm cov}
		+ w_{\rm fp}L^{ik}_{\rm fp} \\
		& + w_{\rm min}L^{k}_{\rm min}
		+ w_{\rm conn}L^{k}_{\rm conn},
	\end{split}
\end{equation}
	where \(w_{\rm sub}\), \(w_{\rm cov}\), \(w_{\rm fp}\), \(w_{\rm min}\), and \(w_{\rm conn}\) are the weights of the corresponding local loss terms.
	
	The subset term is defined as
	\begin{equation}
		L^{ik}_{\rm sub}
		=
		\left\langle
		\max\left(\Gamma_k-M^i_{\rm miss},0\right)
		\right\rangle .
	\end{equation}
	This term penalizes active cells in \(\Gamma_k\) that appear at non-hit positions in \(M^i_{\rm miss}\). It corresponds to the trigger semantics described above: the active positions of the core template should be supported as much as possible by the observed pattern with missing hits, rather than relying on hits that may be absent in the actual observation.
	
	The coverage term is defined as
	\begin{equation}
		L^{ik}_{\rm cov}
		=
		\left\langle
		\max\left(M^i_{\rm clean}-\Delta^k_{\rm soft},0\right)
		\right\rangle .
	\end{equation}
	It requires \(\Delta_{\rm soft}\) to cover the major true hits in the ideal track, thereby preserving the track acceptance efficiency.
	
	The false-positive term is defined as
	\begin{equation}
		L^{ik}_{\rm fp}
		=
		\left\langle
		\Delta^k_{\rm soft}\left(1-M^i_{\rm clean}\right)
		\right\rangle .
	\end{equation}
	It penalizes excessive expansion of \(\Delta_{\rm soft}\) outside the true-track region, thereby reducing the probability of false matches under high-background conditions. Here, \(\langle\cdot\rangle\) denotes the average over all cells in the \(6\times 8\) window.
	
	To prevent the trigger skeleton from becoming overly sparse, a lower-bound constraint on the number of active cells in the trigger skeleton is introduced. Let
	\begin{equation}
	|\Gamma_k|=\sum_{a,b}\Gamma_k(a,b),
	\end{equation}
	then
	\begin{equation}
L_{\rm min}^{k}
=
\frac{\max(4-|\Gamma_k|,0)}{4}.
	\end{equation}
	This term penalizes \(\Gamma\) templates with fewer than four active cells, preventing the network from generating excessively small trigger skeletons.
	
	The connectivity term is used to suppress isolated active points. Let \(\mathcal{N}(a,b)\) denote the eight-neighborhood of cell \((a,b)\), and let the number of active neighboring cells be
	\begin{equation}
	n_k(a,b)
	=
	\sum_{(u,v)\in \mathcal{N}(a,b)}\Gamma_k(u,v).
	\end{equation}
	The connectivity loss is then defined as
	\begin{equation}
	L_{\rm conn}^{k}
	=
	\left\langle
	\Gamma_k(a,b)\exp[-\beta n_k(a,b)]
	\right\rangle_{a,b},
	\end{equation}
	where \(\beta\) is a parameter controlling the penalty strength for isolated points. This term takes a larger value when an active cell lacks neighboring active cells, thereby encouraging \(\Gamma\) to form a more continuous geometric structure.
	
	Since a track pattern can be described by any one of multiple candidate templates, the correspondence between samples and templates is not fixed in advance during training. Instead, a softmin assignment is performed according to the matching cost between each sample and each template:
	\begin{equation}
	q_{ik}
	=
	\frac{\exp(-C_{ik}/\tau)}
	{\sum_{j=1}^{K}\exp(-C_{ij}/\tau)} ,
	\end{equation}
	where \(\tau\) is the temperature parameter. The main loss is defined as
	\begin{equation}
	L_{\rm main}
	=
	\frac{1}{N}
	\sum_{i=1}^{N}
	\sum_{k=1}^{K}
	q_{ik}C_{ik},
	\end{equation}
	where \(N\) denotes the number of samples in the current batch. This mechanism allows different templates to automatically specialize during training and cover different track shapes, while avoiding premature assignment of a sample to a single template in the early training stage.
	
	In addition to \(L_{\rm main}\), several global regularization terms are introduced to control the overall properties of the template library.
	
	First, a template diversity loss is used to suppress high similarity among different \(\Gamma\) templates. The \(k\)-th template is flattened into a vector \(\mathbf{g}_k\), and the normalized similarity is defined as
	\begin{equation}
	s_{kl}
	=
	\frac{\mathbf{g}_k\cdot\mathbf{g}_l}
	{\|\mathbf{g}_k\|\,\|\mathbf{g}_l\|+\epsilon},
	\end{equation}
	where \(\epsilon\) is a small positive constant used to avoid a zero denominator. The diversity term is written as
	\begin{equation}
	L_{\rm div}
	=
	\frac{1}{K(K-1)}
	\sum_{k\ne l}
	s_{kl}^{p},
	\end{equation}
	where \(p\) is a nonlinear enhancement exponent. This term imposes a stronger penalty on highly similar templates, thereby reducing redundant structures in the template library.
	
	The template usage rate is estimated from the soft-assignment weights:
	\begin{equation}
	u_k
	=
	\frac{1}{N}
	\sum_{i=1}^{N}
	q_{ik}.
	\end{equation}
	To estimate whether a template is active, the soft activity is defined as
	\begin{equation}
	a_k
	=
	\sigma\left(
	\frac{u_k-u_{\rm th}}{\sigma_u}
	\right),
	\end{equation}
	where \(u_{\rm th}\) is the usage-rate threshold and \(\sigma_u\) controls the smoothness. The number of soft active templates is
	\begin{equation}
	N_{\rm active}
	=
	\sum_{k=1}^{K}
	a_k .
	\end{equation}
	The inactive-template constraint is defined as
	\begin{equation}
	L_{\rm inactive}
	=
	\max\left(T_{\rm active}-N_{\rm active},0\right)^2,
	\end{equation}
	where \(T_{\rm active}\) is the predefined target number of active templates. This term mainly encourages the number of active templates to be no smaller than the target value, thereby reducing the number of templates that remain unused for a long time.
	
	The usage-balance term is defined as
	\begin{equation}
	L_{\rm bal}
	=
	\sum_{k=1}^{K}
	\left(u_k-\frac{1}{K}\right)^2.
	\end{equation}
	It constrains the usage-rate differences among templates and prevents a small number of templates from accounting for most samples.
	
	The noise-rejection term is constructed using the purely random noise window \(M_{\rm noise}\). For noise sample \(i\) and template \(k\), the coverage ratio of the noise over the trigger skeleton is defined as
	\begin{equation}
	r_{ik}
	=
	\frac{
		\sum_{a,b}\Gamma_k(a,b)M_{\rm noise}^{i}(a,b)
	}
	{|\Gamma_k|+\epsilon}.
	\end{equation}
	When the coverage ratio of pure noise over \(\Gamma_k\) exceeds the threshold \(\rho_{\rm rej}\), it is penalized:
	\begin{equation}
	L_{\rm rej}
	=
	\frac{1}{NK}
	\sum_{i=1}^{N}
	\sum_{k=1}^{K}
	\max\left(r_{ik}-\rho_{\rm rej},0\right).
	\end{equation}
	This term is used to reduce the probability that a template is triggered by random background hits covering the trigger skeleton.
	
	To reduce the discrepancy between the soft template used during training and the exported binary template, a binary-consistency term is introduced. Let
	\begin{equation}
	\Delta_{\rm bin}^{k}
	=
	\mathbf{1}\left(\Delta_{\rm soft}^{k}>\theta_{\Delta}\right),
	\end{equation}
	where \(\theta_{\Delta}\) is the threshold used to export the binary recovery template \(\Delta\). The binary-consistency loss is defined as
%	\begin{equation}
%	L_{\rm bin}
%	=
%	\frac{1}{K}
%	\sum_{k=1}^{K}
%	\left|
%	\Delta_{\rm soft}^{k}
%	-
%	\operatorname{stopgrad}\left(\Delta_{\rm bin}^{k}\right)
%	\right| .
%	\end{equation}
	\begin{equation}
		L_{\rm bin}
		=
		\frac{1}{K}
		\sum_{k=1}^{K}
		\left\langle
		\left|
		\Delta_{\rm soft}^{k}
		-
		\operatorname{stopgrad}\left(\Delta_{\rm bin}^{k}\right)
		\right|
		\right\rangle .
	\end{equation}
	
	Here, \(\operatorname{stopgrad}(\cdot)\) indicates that the binary template is used only as a training calibration target and does not participate in gradient backpropagation. This term reduces the difference between \(\Delta_{\rm soft}\) used in training and the binary \(\Delta\) actually used after export.
	
	In summary, the total loss is written as
	\begin{equation}
	\begin{aligned}
		L ={}& L_{\rm main}
		+\lambda_{\rm div}L_{\rm div}
		+\lambda_{\rm inactive}L_{\rm inactive} \\
		&+\lambda_{\rm bal}L_{\rm bal}
		+\lambda_{\rm rej}L_{\rm rej}
		+\lambda_{\rm bin}L_{\rm bin}.
	\end{aligned}
	\end{equation}
	
	Here, each \(\lambda\) controls the weight of the corresponding global regularization term. Through these constraints, the optimization process balances trigger robustness, recovery coverage, background rejection, and template compactness, thereby avoiding degradation of physics performance caused by optimizing a single metric alone.
	
	It should be noted that the number of templates \(K\) specified at the beginning of training is not necessarily equal to the final number of usable templates. Instead, it represents the candidate-template capacity available for assignment during optimization. Under multi-objective constraints, some candidate templates may receive very low soft-assignment weights for a long time and appear as inactive templates at the export stage. This work allows the existence of such inactive templates, because forcing all candidate templates to become active may increase redundancy among templates or cause some templates to deviate from physically reasonable hit structures merely to obtain usage. In the current implementation, a single training run produces a set of effective templates under a given template budget and a given training-sample distribution, rather than a complete template library that fully covers all relevant local track segments. For a wider momentum or background range, the phase space can be divided into several finer subranges and trained separately, or multiple training runs can be performed. The effective templates exported from these runs can then be merged into the final template library through a unified export procedure. The export process is a standardized consolidation of the learned candidate templates, rather than a manual redesign of the template structures. The main structures of the templates are still learned automatically through differentiable optimization.
	
	% ============================================================
	\section{Algorithm Performance}
	\label{sec:algorithm_performance}
	
	The pattern matching algorithm can be divided into two parts: the pattern representation and the matching procedure. This work mainly evaluates the physics performance of the original matrix-based pattern matching algorithm after deploying the \(\Gamma/\Delta\) binary template pairs constructed and exported by the neural-network-assisted framework. Optimization on general-purpose processors for improving algorithm speed and throughput is not the main focus of this study. Nevertheless, at the end of this section, we discuss the relationship between the number of templates and the throughput, so that researchers can make trade-offs between physics performance and processing speed.
	
	\subsection{Data Samples}
	\label{subsec:data_samples}
	
	The Offline Software of Super Tau-Charm Facility (OSCAR)~\cite{OSCAR1,OSCAR2} is mainly used for offline data processing, including detector simulation, digitization, reconstruction, and physics analysis. Using OSCAR, we generated five physics-process samples with charged final states. The covered energy points include the \(J/\psi\) peak at \(3.097~\mathrm{GeV}\), as well as \(4.26~\mathrm{GeV}\) and \(4.682~\mathrm{GeV}\). The momentum range of the final-state particles is approximately \([0.1,2]~\mathrm{GeV}/c\), covering both low- and high-momentum regions. The spatial distribution covers the polar-angle range of \([20^\circ,160^\circ]\), corresponding to the detector acceptance.

	In the experiments where background samples are mixed to simulate realistic conditions, complete events are used and mixed with different background levels. The ratios of background hits to signal hits under different background levels are listed in Table~\ref{tab:hit_occupancy}. In the following, we denote the background level by \(N{\rm Bkg}\) with \(N=1,2,3\). The backgrounds mainly include beam-related backgrounds and luminosity-related backgrounds. Under the current accelerator design, their ratio is approximately \(6:1\)~\cite{OSCAR3}. When testing the signal retention ratio of the algorithm in different momentum intervals and under different detector-efficiency conditions, we mainly use single-particle muon events. After backgrounds are introduced, background hits may trigger additional templates, causing part of the true hits to be retained indirectly and thereby increasing the signal retention ratio. However, this increase does not indicate an improvement in pattern matching quality. Instead, it also introduces more background hits, increases the output fraction, and thus degrades the data-compression performance. Therefore, when evaluating the coverage capability of templates for true track shapes, this work uses single-particle muon samples without background mixing. Unless explicitly stated as tests using single-particle samples, the results associated with specific physics-channel names in the following refer to complete event samples mixed with the corresponding background level.

	\begin{table*}[t]
		\centering
		\caption{MDC hit occupancy statistics for five physics channels under different background levels. The real-hit ratio is defined as the ratio of hits from true particle tracks to all hits. The hit occupancy ratio is defined as the ratio of the number of hit sense wires in each event to the total number of sense wires. The values are averaged over \(10^4\) events.}
		\label{tab:hit_occupancy}
		\resizebox{\textwidth}{!}{
			\begin{tabular}{lcccccc}
				\toprule
				\multirow{2}{*}{Physics channel}
				& \multicolumn{2}{c}{\(1{\rm Bkg}\)}
				& \multicolumn{2}{c}{\(2{\rm Bkg}\)}
				& \multicolumn{2}{c}{\(3{\rm Bkg}\)} \\
				\cmidrule(lr){2-3}
				\cmidrule(lr){4-5}
				\cmidrule(lr){6-7}
				& Real-hit ratio & Hit occupancy ratio
				& Real-hit ratio & Hit occupancy ratio
				& Real-hit ratio & Hit occupancy ratio \\
				\midrule
				\makecell[l]{\(e^+e^- \to \pi^+\pi^- J/\psi\)\\ \(J/\psi \to e^+e^-\)}
				& 0.270 & 0.066
				& 0.161 & 0.111
				& 0.117 & 0.153 \\
				
				\makecell[l]{\(e^+e^- \to \pi^+\pi^- J/\psi\)\\ \(J/\psi \to \mu^+\mu^-\)}
				& 0.264 & 0.067
				& 0.159 & 0.111
				& 0.115 & 0.153 \\
				
				\makecell[l]{\(e^+e^- \to K^+K^- J/\psi\)\\ \(J/\psi \to l^+l^-\)}
				& 0.303 & 0.070
				& 0.184 & 0.115
				& 0.136 & 0.156 \\
				
				\(e^+e^- \to \tau^+\tau^-\)
				& 0.156 & 0.058
				& 0.087 & 0.103
				& 0.062 & 0.146 \\
				
				\(J/\psi \to \Lambda\bar{\Lambda}\)
				& 0.344 & 0.074
				& 0.214 & 0.118
				& 0.158 & 0.160 \\
				\bottomrule
			\end{tabular}
		}
	\end{table*}

	\subsection{Metric Definitions}
	\label{subsec:metric_definitions}
	
	The HLT system needs to reject invalid or background hits while preserving true physics signals as much as possible, thereby reducing the pressure on backend storage and the computational load of subsequent reconstruction. Therefore, this work mainly uses the signal retention ratio, output fraction, background rejection rate, and hit-on-track efficiency to evaluate the physics performance of the template library.
	
	The signal retention ratio is defined as the ratio of the number of true track hits retained by the algorithm to the number of input true track hits:
	\begin{equation}
	\epsilon_{\rm sig}
	=
	\frac{N_{\rm sig}^{\rm out}}{N_{\rm sig}^{\rm in}} .
	\end{equation}
	
	The output fraction is defined as the ratio of the total number of output hits to the total number of input hits:
	\begin{equation}
	f_{\rm out}
	=
	\frac{N_{\rm hit}^{\rm out}}{N_{\rm hit}^{\rm in}} .
	\end{equation}
	
	The background rejection rate is defined as the ratio of background hits not retained by the algorithm to the number of input background hits:
	\begin{equation}
	r_{\rm bkg}
	=
	1 -
	\frac{N_{\rm bkg}^{\rm out}}{N_{\rm bkg}^{\rm in}} .
	\end{equation}
	
	The hit-on-track efficiency is defined as the ratio of the number of hits on a given true track retained by the algorithm to the total number of input hits belonging to that track.
	
	When detector efficiency \(\eta\) is considered, finite detector efficiency is simulated by randomly dropping a fraction \((1-\eta)\) of both true track hits and background hits. For the metrics above, the denominators are defined using the hits that actually enter the algorithm after the detector-efficiency effect is applied. Therefore, the signal retention ratio, output fraction, and background rejection rate all lie between 0 and 1.
	
	\subsection{Physics Performance}
	\label{subsec:physics_performance}

	This work focuses on template construction and local hit retention in matrix-based pattern matching at the local segment finding stage. This step is performed before the complete MDC two-dimensional track reconstruction and provides candidate local segments for subsequent algorithms. It obtains candidate local segments while rejecting noise-hit clusters that clearly do not conform to track-like shapes. The segment combination algorithm then combines these candidate local segments into complete tracks and rejects false candidates. The reconstruction algorithm then completes the MDC two-dimensional reconstruction at the HLT stage based on the tracks formed from these candidate local segments. Therefore, for the local segment finding algorithm at this stage, the physics-performance study mainly focuses on the relationship among the number of template pairs, the covered transverse-momentum range, detector efficiency, and signal retention ratio. Since the input hits of the segment combination and reconstruction algorithms come entirely from the local segment finding algorithm, a high retention rate is required at this stage; otherwise, inefficiencies will propagate to all subsequent algorithms. In contrast, background hits can be further rejected in the segment combination and reconstruction stages using global information across superlayers. Therefore, at this stage, while ensuring a high retention rate, the algorithm rejects hit clusters that are clearly inconsistent with track-like shapes. For this reason, the final track or segment reconstruction efficiency is not used as the primary evaluation metric in this work. Instead, hit-level signal retention, hit-on-track efficiency, background rejection, and output fraction are used to evaluate the performance of the template library at this stage.

\begin{figure}[tbp]
		\centering
		\includegraphics[width=0.80\columnwidth]{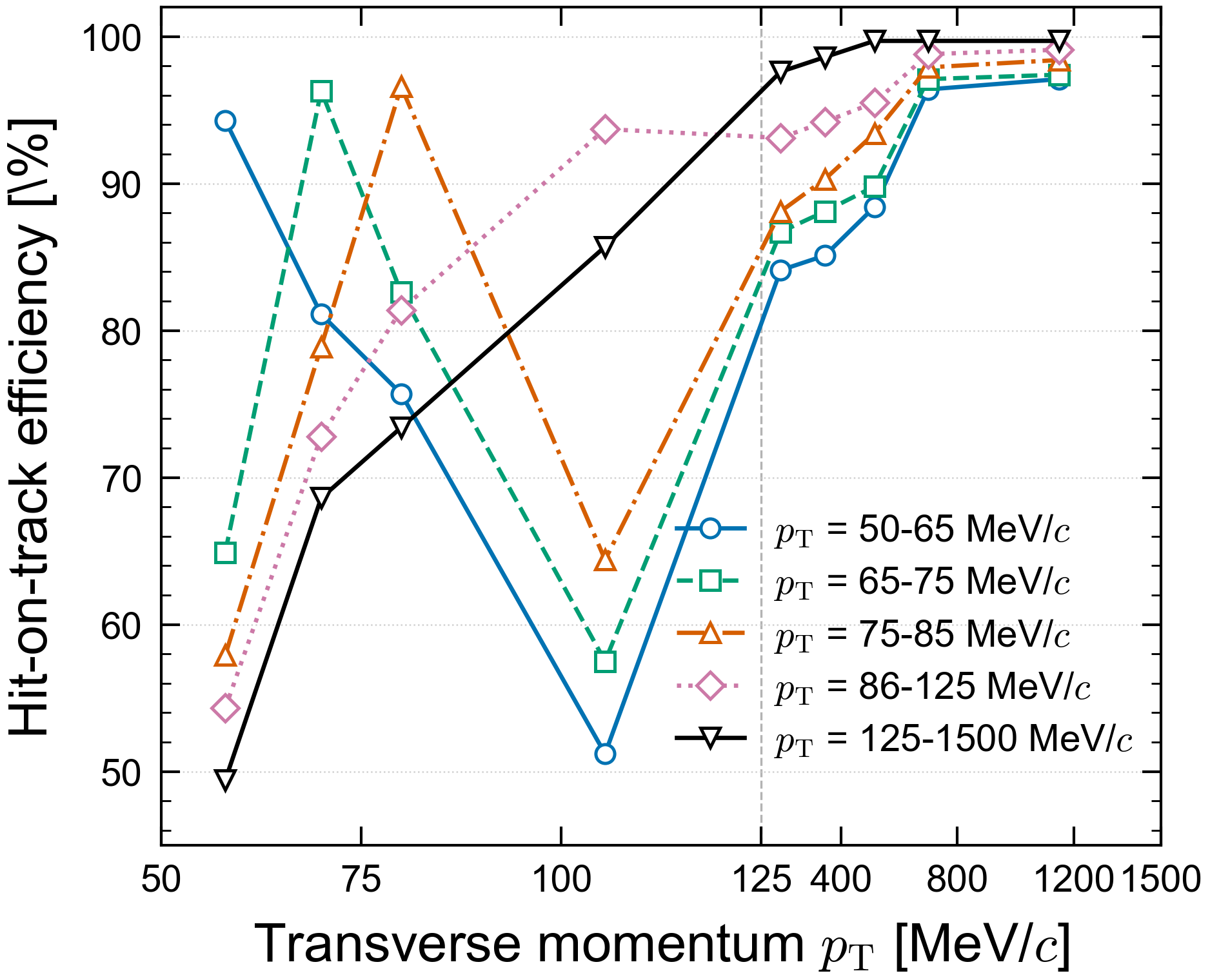}
		\caption{Hit-on-track efficiency in the momentum range of \(50\)--\(1500~\mathrm{MeV}/c\) for template libraries trained with datasets in different transverse-momentum intervals. The transverse-momentum intervals used for template-library construction are divided into five groups according to track morphology. }
		\label{fig:retention_momentum_range}
	\end{figure}

	Single-particle events are used to evaluate the hit-on-track efficiency of different template libraries.  Fig.~\ref{fig:retention_momentum_range} shows the relationship between hit-on-track efficiency and template libraries trained using datasets covering different transverse-momentum intervals. As discussed above, the momentum coverage of a template library is determined by the \(p_T\) interval of the training samples. We can manually specify the momentum range of the particle tracks in the input dataset to generate templates for a particular interval. Owing to the geometrical relationship among MDC superlayers, for the same particle track, the local segment left in an outer superlayer is more inclined than that in an inner superlayer. Therefore, more complex particle local segments tend to appear first in the outermost superlayers. The momentum grouping is mainly performed according to the local segment in the outermost superlayer that the particle can reach. As a result, even if two momentum groups do not overlap, their local segments in the inner superlayers can still be similar. Therefore, a template library trained using tracks in a certain momentum interval can cover a working range broader than that interval itself, mainly extending toward higher-momentum regions. Since the sliding window itself does not require the particle to traverse the entire superlayer completely, but only needs to capture the local segment features and compare them with the trigger template \(\Gamma\), the algorithm can still find local segments for medium- and low-momentum tracks.
	
	To study the template-pair count and performance in different transverse-momentum intervals when the template coverage is extended from medium--high momentum to medium--low momentum, the template pairs used for gradual extension are divided into five groups in this work. The first four groups are constructed by the neural-network-assisted framework according to the morphology of local segments in the sliding window. They correspond to cases where the track points in the \(\Gamma\) template are mainly distributed in columns 1--2, columns 1--4, columns 1--8, and shapes that do not pass through all rows of a single superlayer but mainly extend along the \(\varphi\) direction, respectively. These four template groups roughly correspond to high-momentum, medium--high-momentum, medium--low-momentum, and low-momentum to very-low-momentum or large-curvature local segments. The fifth group is a manually supplemented template group based on local-segment morphology and occurrence frequency. It is mainly used for comparison with automatically constructed templates and for evaluating the effect of further extending low-momentum coverage on the hit-on-track efficiency.

\begin{figure}[tbp]
		\centering
		\includegraphics[width=0.80\columnwidth]{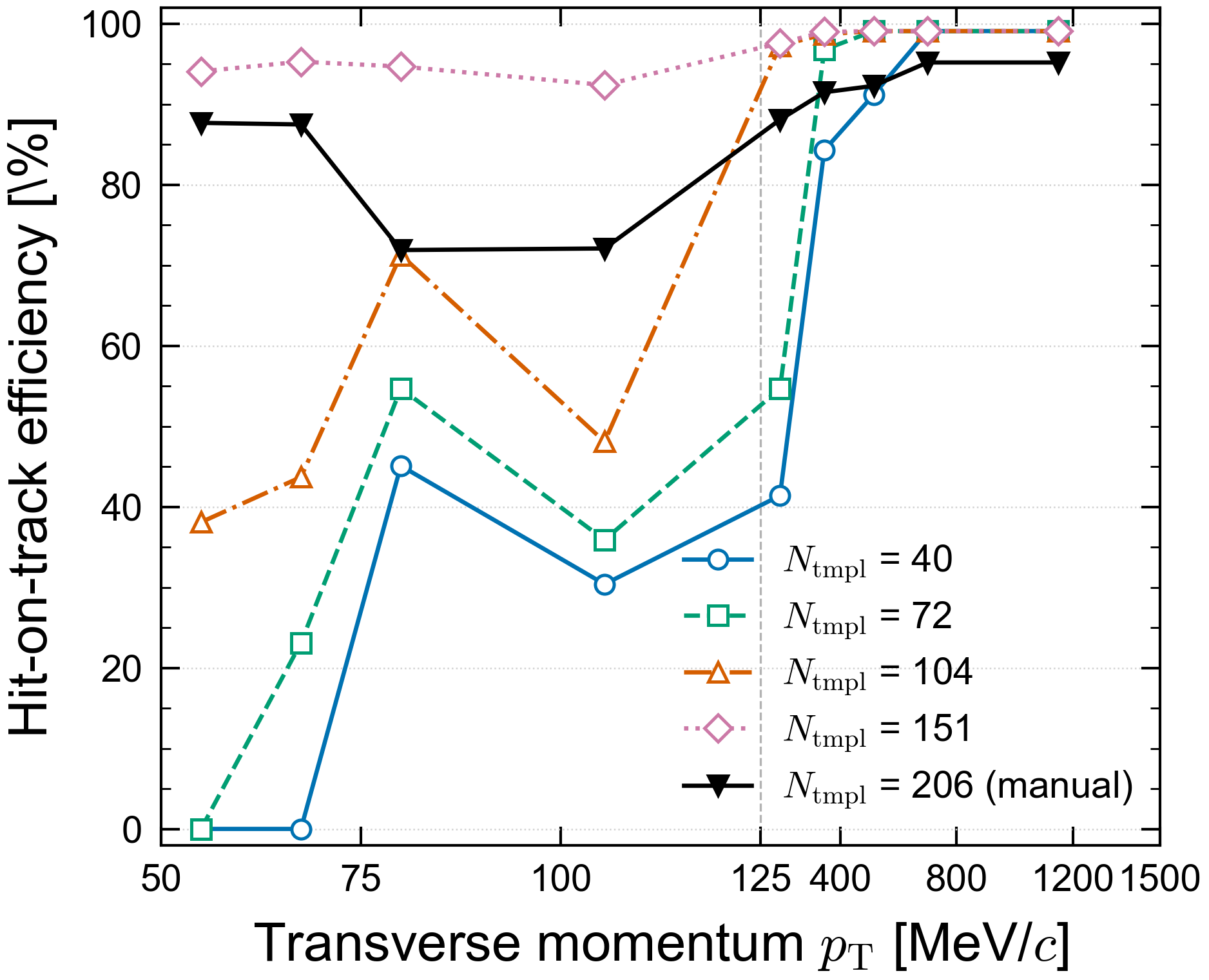}
		\caption{Relationship between the number of template pairs and the hit-on-track efficiency in different momentum intervals after progressively extending the template library. The detector efficiency is set to \(95\%\). As the number of templates increases, the hit-on-track efficiency in the medium--low-momentum intervals improves.}
		
		\label{fig:template_number_retention}
	\end{figure}

	Fig.~\ref{fig:template_number_retention} shows the relationship between the number of template pairs and the hit-on-track efficiency in different transverse-momentum intervals as the above template groups are added sequentially. In the figure, an efficiency decrease is observed in the \(110\)--\(130~\mathrm{MeV}/c\) interval. This is mainly because, in this momentum range, the particle-track radius is close to the outer radius of the MDC. Consequently, the local segments in the outermost superlayers are significantly elongated along the \(\varphi\) direction and may exceed the fixed window size. Meanwhile, in template optimization for environments with background hits, the sparse \(\Gamma\) templates corresponding to such local segments are more likely to be falsely triggered by background hits. Therefore, under a limited template budget, the optimization process tends to prioritize more stable and more common local-segment morphologies.
	
\begin{figure}[tbp]
		\centering
		\includegraphics[width=0.80\columnwidth]{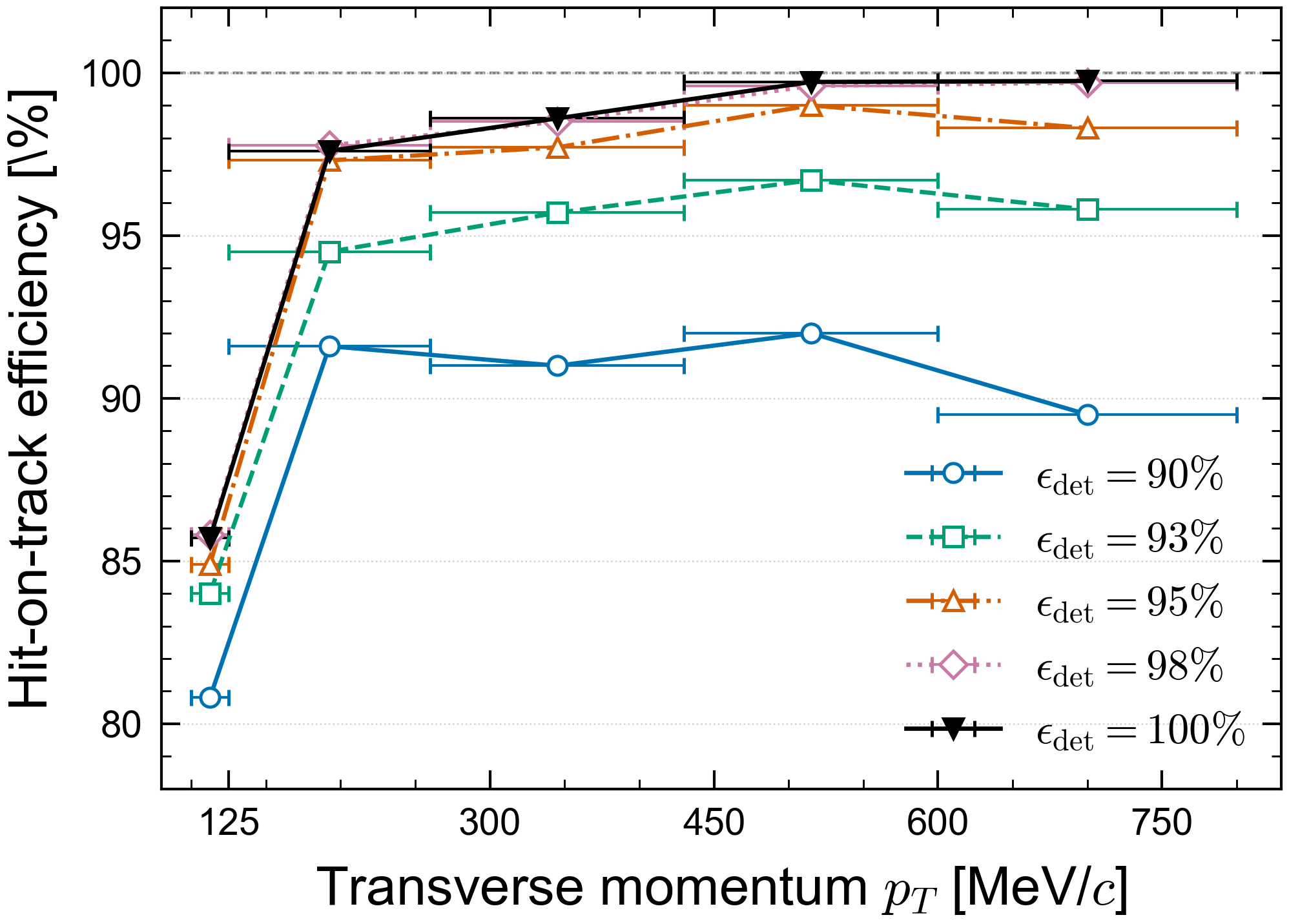}
		\caption{Hit-on-track efficiency as a function of detector efficiency in \(100\)--\(800~\mathrm{MeV}/c\) single-particle samples. The template library is constructed using training samples in the \(120\)--\(1500~\mathrm{MeV}/c\) range, with a preset detector efficiency of \(95\%\) during training. It contains 104 template pairs. When the actual detector efficiency is higher than or close to the training setting, the template library maintains a high hit-retention capability. When the actual detector efficiency is lower than the training setting, the retention efficiency decreases moderately.}
		\label{fig:detector_efficiency_hot}
	\end{figure}

	Fig.~\ref{fig:detector_efficiency_hot} shows the hit-on-track efficiency in single-particle samples under different detector efficiencies and different \(p_T\) intervals. The transverse-momentum range of the particles used to construct the template library is \(120\)--\(1500~\MeVc\), as explained later. Since the local segments formed by hits at the sense-wire matrix level are already approximately straight when the particle transverse momentum exceeds \(800~\MeVc\), the result in the \(600\)--\(800~\MeVc\) interval provides a conservative indication of the retention behavior for tracks with \(p_T>800~\MeVc\), whose local segments are generally no more curved at the sense-wire matrix level.

	After constructing binary templates with the neural-network-assisted framework in the above study, local segments from particle tracks with momenta above \(50~\mathrm{MeV}/c\) can be retained, showing the coverage of neural-network-assisted template construction. However, in the current engineering implementation, matrix-based pattern matching is used for local segment finding with \(p_T>120~\mathrm{MeV}/c\). This choice is based on two considerations.

	As the particle transverse momentum decreases further, the \(\Gamma\) templates trained for the medium--low-\(p_T\) region and for local segments distributed along the \(\varphi\) direction tend to become sparser under reduced detector-efficiency conditions. Under multiple-background conditions, this leads to an increase in the output fraction, thereby reducing the data-compression performance. Table~\ref{tab:output_fraction_template_number} uses the \(e^+e^- \to \pi^+\pi^- J/\psi\),\(J/\psi \to e^+e^-\) channel as an example to show how the output fraction changes as the number of template pairs increases and the coverage is further extended toward lower transverse momentum.
	
	From the perspective of system engineering, low-momentum tracks that do not fully traverse the MDC require more complex segment combination algorithms. For such tracks, even under high-background conditions, the track features are relatively distinctive, and algorithms such as the Hough transform can directly obtain the track representation. Therefore, in the subsequent evaluation, the working transverse-momentum range of the template library is set to \(p_T>120~\mathrm{MeV}/c\).

	\begin{table*}[t]
		\centering
		\caption{Output fraction of template libraries with different numbers of template pairs under different background levels in the \(\pi^+\pi^-J/\psi, J/\psi\to e^+e^-\) channel.}
		\label{tab:output_fraction_template_number}
		\begin{tabular}{c|ccccc}
			\toprule
			Background level & 40 & 72 & 104 & 151 & 206 (manual) \\
			\midrule
			\(1{\rm Bkg}\) & 0.358 & 0.437 & 0.522 & 0.594 & 0.514 \\
			\(2{\rm Bkg}\) & 0.401 & 0.482 & 0.536 & 0.608 & 0.519 \\
			\(3{\rm Bkg}\) & 0.453 & 0.500 & 0.577 & 0.651 & 0.557 \\
			\bottomrule
		\end{tabular}
	\end{table*}

	In Table~\ref{tab:output_fraction_template_number}, 40, 72, 104, and 151 denote the numbers of template pairs in the neural-network-assisted template libraries obtained by progressively extending the coverage from high momentum to low momentum. The entry ``206 (manual)'' denotes the conventional manually constructed template library. All results are obtained from complete event samples of the \(e^+e^- \to \pi^+\pi^- J/\psi\),\(J/\psi \to e^+e^-\) channel mixed with \(1\), \(2\), and \(3\) times the nominal background level, respectively. The detector efficiency is set to \(95\%\). The output fraction is defined as the ratio of the total number of output hits to the total number of input hits. A larger value indicates weaker data-compression performance.
	
	Although the manually constructed template library gives a lower output fraction, Fig.~\ref{fig:template_number_retention} shows that its signal retention ratio is also relatively low. It is worth noting that, after templates covering the medium--low-momentum region are added, the signal retention ratio increases rapidly, while the output fraction also rises. This indicates that the additional templates improve the coverage of true tracks, but may also introduce more false triggers and retain extra hits. A counterintuitive observation is that the newly added medium--low-momentum templates correspond to local segments with larger inclination angles. In principle, such local segments have more distinctive features and should be less likely to be confused with random background hits; therefore, they would not be expected to substantially increase the false-trigger rate. Further analysis shows that inclined tracks are more sensitive to detector inefficiency. Under the joint loss constraints, in order to maintain trigger robustness under reduced detector-efficiency conditions, the optimization process tends to learn sparser \(\Gamma\) trigger skeletons, which are more easily triggered by dense background hits. Moreover, because the \(\Delta\) recovery regions corresponding to inclined tracks are usually larger, a single false trigger may retain more hits.

	Therefore, in practical HLT applications, a trade-off is made between the momentum coverage of the template library and the output fraction. If the template library extensively covers the medium--low-momentum region, the output fraction will increase, and the number of candidate local segments will also increase. In the current prototype study, matrix-based pattern matching is used for local segment finding with \(p_T>120~\mathrm{MeV}/c\). For lower-momentum tracks that do not fully traverse the MDC, global track-finding algorithms such as the two-dimensional Hough transform can be considered to directly estimate potential track parameters. The low-momentum template library can still serve as a preselection or denoising tool, but its joint optimization with subsequent global track-finding algorithms is a system-level engineering problem and will be investigated in future work.
	
	Table~\ref{tab:template_method_comparison} compares the evaluation results of the neural-network-assisted binary templates, the conventional manually constructed templates, and the templates constructed from frequency statistics. All results in the table are obtained in the \(e^+e^- \to \pi^+\pi^- J/\psi\), \(J/\psi \to e^+e^-\) channel under the \(1{\rm Bkg}\) condition and the specified detector efficiency. TOP1000 denotes the template library constructed by selecting the 1000 most frequent local segment patterns from the training samples under the corresponding detector efficiency. In this case, \(\Gamma=\Delta\) is imposed to exclude the possibility that the \(\Delta\) recovery region additionally introduces non-track hits. Manual denotes the template library obtained through manual construction and selection. Manual (90\%) denotes a template library constructed on top of the manual templates by randomly removing \(10\%\) of the hits to form incomplete templates, with the aim of enhancing robustness against missing hits. NN denotes the template library exported by the neural-network-assisted binary template construction framework.
	
	Taking the result of the manually constructed templates at \(100\%\) detector efficiency as a reference, one can see that, when the detector efficiency is set to \(95\%\), both the frequency-based and manually constructed templates suffer from insufficient robustness. In contrast, the templates constructed using the neural-network-assisted framework achieve, at \(95\%\) detector efficiency, a signal retention ratio close to that of the manually constructed templates under the ideal \(100\%\) detector-efficiency condition. This demonstrates the effectiveness of the proposed optimization.
	
	The table also shows that the background rejection rates of different template construction methods are limited. Further inspection indicates that this phenomenon is not mainly caused by the \(\Delta\) recovery regions. Instead, it arises because, at the local segment finding stage, the algorithm does not use global association information across superlayers. As a result, some candidate fragments formed by background hits but exhibiting local segment-like features within a single superlayer are still retained. By comparing TOP1000 at \(100\%\) detector efficiency with the NN-based templates at \(95\%\) detector efficiency, it can be seen that, when the signal retention ratios are close, the background rejection rate of the NN-based templates is only about \(5.2\) percentage points lower than that of the TOP1000 templates. This indicates that the limited background rejection rate is not mainly caused by the recovery behavior of the \(\Delta\) templates. Rather, it is because isolated candidate local segments produced by background hits are not rejected at this stage, and their hits are consequently retained.
	
	These retained background-induced local segments can be further rejected by the segment combination algorithm using global information, thereby reducing the output fraction in subsequent processing. At the local segment finding stage, these background-induced local segment signals are therefore retained. Therefore, as the first step of track reconstruction, local segment finding places emphasis on achieving a high signal retention ratio.

	\begin{table*}[t]
		\centering
		\caption{Comparison of different template construction methods in the \(e^+e^- \to \pi^+\pi^- J/\psi\),\(J/\psi \to e^+e^-\) channel under the \(1{\rm Bkg}\) condition.}
		\label{tab:template_method_comparison}
		\resizebox{\textwidth}{!}{
			\begin{tabular}{lccccc}
				\toprule
				Method & Detector efficiency & Number of template pairs & Signal retention ratio & Output fraction & Background rejection \\
				\midrule
				Manual & \(100\%\) & 80 & 0.984 & 0.447 & 0.751 \\
				Manual & \(95\%\) & 80 & 0.742 & 0.394 & 0.734 \\
				Manual (90\%) & \(95\%\) & 448 & 0.861 & 0.476 & 0.666 \\
				TOP1000 & \(95\%\) & 1000 & 0.897 & 0.482 & 0.671 \\
				TOP1000 & \(100\%\) & 1000 & 0.967 & 0.483 & 0.699 \\
				NN & \(95\%\) & 104 & 0.979 & 0.522 & 0.647 \\
				\bottomrule
			\end{tabular}
		}
	\end{table*}
	
	Using the current binary templates constructed with the neural-network-assisted framework, the algorithm performance in five channels under different background levels is shown in Table~\ref{tab:five_channel_results}. 	The results in the table are obtained using 104 neural-network-assisted \(\Gamma/\Delta\) template pairs. This template library is constructed from training samples with \(p_T=120\)--\(1500~\mathrm{MeV}/c\). During testing, the detector efficiency is set to \(95\%\), and complete events from each physics channel are mixed with \(1\), \(2\), and \(3\) times the nominal background level, respectively. The results demonstrate that the algorithm maintains a high signal retention ratio even under multiple-background conditions, while the increase in output fraction remains limited.
	
	\begin{table*}[t]
		\centering
		\caption{Signal retention ratio and output fraction of the neural-network-assisted template library in five physics channels under different background levels.}
		\label{tab:five_channel_results}
		\resizebox{\textwidth}{!}{
			\begin{tabular}{lcccccc}
				\toprule
				\multirow{2}{*}{Physics channel}
				& \multicolumn{2}{c}{\(1{\rm Bkg}\)}
				& \multicolumn{2}{c}{\(2{\rm Bkg}\)}
				& \multicolumn{2}{c}{\(3{\rm Bkg}\)} \\
				\cmidrule(lr){2-3}
				\cmidrule(lr){4-5}
				\cmidrule(lr){6-7}
				& Output fraction & Signal retention ratio
				& Output fraction & Signal retention ratio
				& Output fraction & Signal retention ratio \\
				\midrule
				\makecell[l]{\(e^+e^- \to \pi^+\pi^- J/\psi\)\\ \(J/\psi \to e^+e^-\)}
				& 0.522 & 0.979
				& 0.536 & 0.983
				& 0.577 & 0.988 \\
				
				\makecell[l]{\(e^+e^- \to \pi^+\pi^- J/\psi\)\\ \(J/\psi \to \mu^+\mu^-\)}
				& 0.523 & 0.978
				& 0.528 & 0.983
				& 0.563 & 0.989 \\
				
				\makecell[l]{\(e^+e^- \to K^+K^- J/\psi\)\\ \(J/\psi \to l^+l^-\)}
				& 0.493 & 0.962
				& 0.501 & 0.975
				& 0.524 & 0.979 \\
				
				\(e^+e^- \to \tau^+\tau^-\)
				& 0.437 & 0.972
				& 0.481 & 0.982
				& 0.504 & 0.987 \\
				
				\(J/\psi \to \Lambda\bar{\Lambda}\)
				& 0.553 & 0.954
				& 0.557 & 0.970
				& 0.579 & 0.974 \\
				\bottomrule
			\end{tabular}
		}
	\end{table*}

	\subsection{GPU Time Consumption and Throughput}
	\label{subsec:gpu_time}

\begin{figure}[tbp]
		\centering
		\includegraphics[width=0.99\columnwidth]{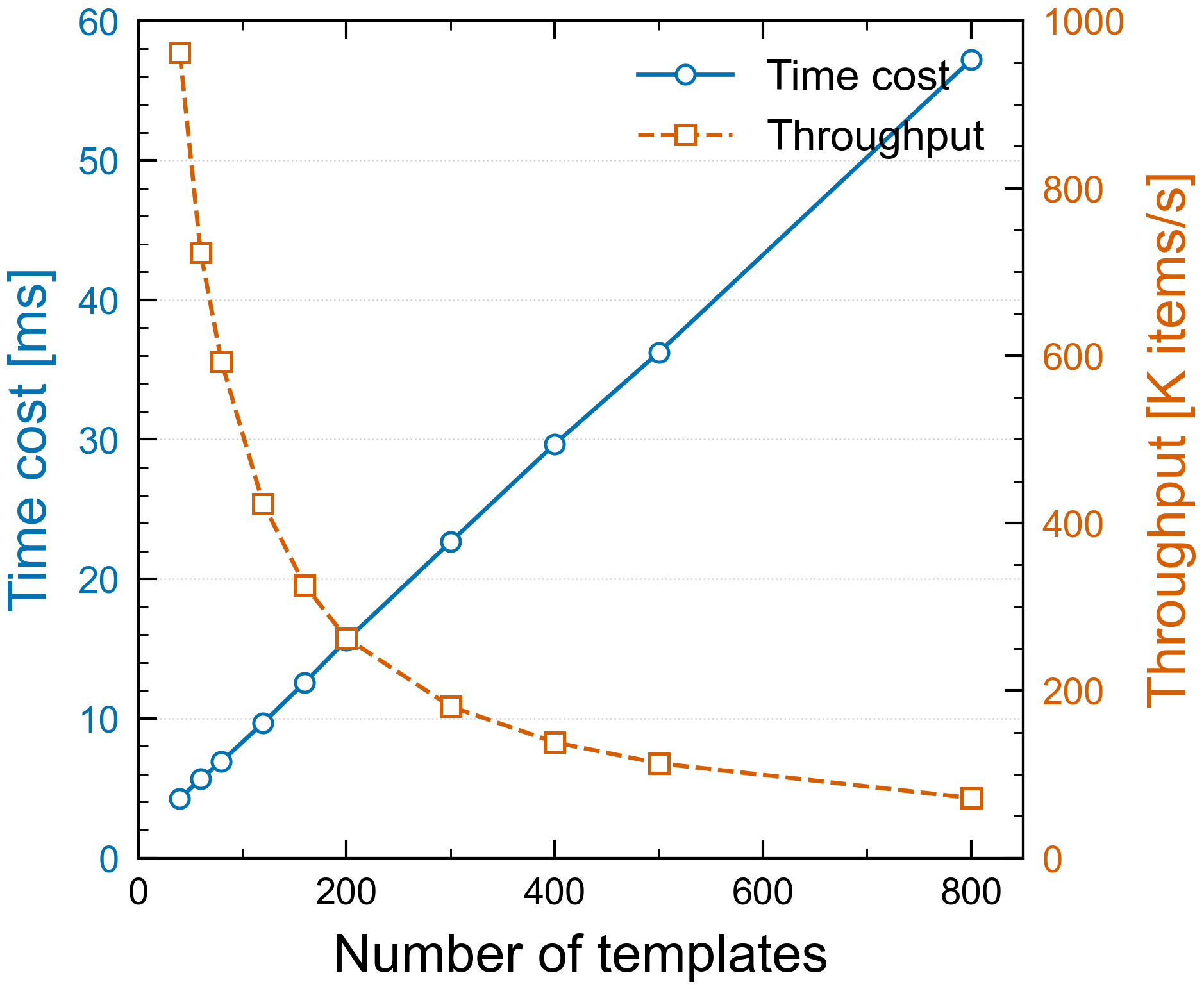}
		\caption{GPU processing time for a fixed batch size of 1024 on a single GPU as a function of the number of templates, together with the corresponding estimated full-card throughput. The measurements are performed on a Tesla A16 computing card with four independent GPUs; the full-card throughput is estimated by processing different batches in parallel on the four GPUs.}

		\label{fig:gpu_throughput}
	\end{figure}

	As discussed above, pattern matching can be divided into two parts: the template representation and the matching procedure. After the template library is constructed using the neural-network-assisted framework, the online stage only needs to load the exported binary templates into the matrix-based pattern matching algorithm and perform bitwise matching operations. Fig.~\ref{fig:gpu_throughput} shows the GPU processing time for a fixed batch size of 1024 on a single GPU and the corresponding estimated full-card throughput as the number of templates increases.

%	The test results show that, when the number of templates is approximately 200 pairs, the batch-averaged processing time on a single GPU is about \(16~\us/\mathrm{event}\), and the throughput of the entire computing card exceeds \(200\,\mathrm{k}\) events/s. Since different events are independent of each other, the HLT online system can process different batches of events in parallel using multiple GPUs or multiple cards. These results indicate that the matrix-based pattern matching stage has good parallel scalability and strong potential for online deployment~\cite{SIMT,CUDA}.

	The results show that, when the number of templates is approximately 200 pairs, the processing time for one batch of 1024 events on a single GPU is about \(16~\mathrm{ms}\), corresponding to a batch-averaged processing time of about \(16~\us/\mathrm{event}\) and a single-GPU throughput of about \(60\,\mathrm{k}\) events/s. Since the Tesla A16 computing card contains four independent GPUs, the estimated full-card throughput can exceed \(200\,\mathrm{k}\) events/s when different batches are processed in parallel on the four GPUs. Since different events are independent of each other, the HLT online system can process different batches of events in parallel using multiple GPUs or multiple cards~\cite{SIMT,CUDA}. These results indicate that the matrix-based pattern matching stage can be efficiently parallelized and has the potential for online deployment.

%	This method decouples template construction from online pattern matching. Physics researchers can obtain the desired physics performance by constructing and optimizing the templates, whereas the online computing performance is mainly determined by the number of templates and the hardware implementation. Therefore, once the number of templates in the template library is fixed, the computational cost of the matrix-based pattern matching stage mainly varies with the template count and is only weakly dependent on the specific geometrical shape of each individual template. As a result, physics-performance optimization and online-computing optimization can be carried out relatively independently.
%	
	This method separates template construction from online pattern matching. Physics researchers can obtain the desired local-segment-finding performance by constructing and optimizing the templates, whereas the online computing performance is mainly determined by the number of templates and the hardware implementation. Therefore, once the number of templates in the template library is fixed, the computational cost of the matrix-based pattern matching stage is mainly determined by the template count and the hardware implementation, and is only weakly dependent on the specific geometrical shape of each individual template. As a result, local-segment-finding performance optimization and online-computing optimization can be carried out relatively independently.

	% Fig. 5 is single-column
	
	% ============================================================
	\section{Conclusion}
	\label{sec:conclusion}
	
	This paper proposes a neural-network-assisted binary template construction framework to improve MDC local segment finding and associated hit retention based on matrix-based pattern matching. In this work, the neural network is used as a differentiable optimization tool to learn template structures satisfying the physical semantics of \(\Gamma/\Delta\). After training, only the binary templates are exported and deployed in the original matrix-based pattern matching algorithm.
	
	The experimental results show that, compared with template libraries constructed using conventional rule-based methods, the proposed method retains more true track hits in the medium--low-transverse-momentum region and enhances the robustness of the template library under non-ideal detector-efficiency conditions. In single-particle tests, the extended template library can cover charged-particle local segments with transverse momenta of approximately \(50\)--\(1500~\mathrm{MeV}/c\). In the main working region of \(p_T>120~\mathrm{MeV}/c\), the template library still maintains a high hit-on-track efficiency at a detector efficiency of \(95\%\). Under the tested \(1\)--\(3\) times background conditions, the method achieves a high signal retention ratio while keeping the output fraction within an acceptable range. When matrix-based pattern matching is executed on a GPU, the average processing time per event remains at the level of several tens of \(\us\), demonstrating its potential for online deployment.
	
	An important feature of this work is that the neural network does not replace the fast trigger or fast reconstruction algorithm itself, but serves as an offline optimization tool for template library construction. In this framework, the conventional bitwise pattern-matching algorithm handles the online decision-making part that is physically constrained, interpretable, and structurally explicit; the neural network handles the high-dimensional nonlinear combinatorial optimization that is difficult to model through manual rules — transforming interrelated physics requirements such as local-segment coverage, robustness to detector inefficiency, background suppression, and template compactness into differentiable constraints, and automatically adjusting the template structure through a joint loss function without the need to manually design complex combinatorial rules. The resulting templates retain clear physical semantics, can be directly binarized, and deployed into the original matrix-based pattern matching algorithm, with the online execution stage involving only deterministic bitwise operations. In addition to pattern matching at the HLT stage, this type of method may also provide useful guidance for trigger-template design at the L1 stage and other fast track-recognition scenarios that rely on template matching.
	
	In future work, we plan to optimize the neural-network architecture and its parameters to improve the coverage of long-tail local segments formed when the track radius is close to the outer radius of the MDC. In addition, we are conducting a systematic evaluation to increase the number of templates within the available resource budget and further reduce false triggers from background hits under low detector-efficiency conditions. Since multi-wire drift chambers and similar gaseous tracking detectors are widely used in high-energy physics experiments, the neural-network-assisted binary template construction strategy proposed in this work is also expected to be transferable to other fast track-recognition scenarios that rely on template matching.
	
\section*{Acknowledgements}

This work is supported by the National Natural Science Foundation of China (No. 12341503); the Hunan Provincial Natural Science Foundation (No.2023RC4006); the Scientific Research Fund of Hunan Provincial Education Department (No.24B0454); the National Key R\&D Program of China (No. 2022YFA1602200 and No. 2023YFA1607200); the international partnership program of the Chinese Academy of Sciences (No. 211134KYSB20200057). We thank the Hefei Comprehensive National Science Center for their strong support on the STCF key technology research project.


\begin{thebibliography}{99}
		
		\bibitem{STCFdesign}
		M. Achasov et al., Front. Phys. 19, 14701 (2024).
		\url{https://doi.org/10.1007/s11467-023-1333-z}
		
		\bibitem{ATLASTDAQ}
		G. Avolio et al., Phys. Procedia 37, 1819--1826 (2012).
		\url{https://doi.org/10.1016/j.phpro.2012.03.755}
		
		\bibitem{BESIII}
		BESIII Collaboration, Nucl. Instrum. Methods Phys. Res. A 598, 7--11 (2009).
		\url{https://doi.org/10.1016/j.nima.2008.08.072}
		
		\bibitem{BelleIIDAQ}
		S. Yamada et al., IEEE Trans. Nucl. Sci. 62, 1175--1180 (2015).
		\url{https://doi.org/10.1109/TNS.2015.2424717}
		
		\bibitem{ATLASHLT}
		M. Nessi et al., ATLAS high-level trigger, data-acquisition and controls:
		Technical Design Report, ATLAS-TDR-016, CERN-LHCC-2003-022 (2003).
		
		\bibitem{ATLASHLT2}
		N. Berger et al., J. Phys. Conf. Ser. 119, 022013 (2008).
		\url{https://doi.org/10.1088/1742-6596/119/2/022013}
		
		\bibitem{CMSHLT}
		A. Hayrapetyan et al., J. Instrum. 19, P11021 (2024).
		\url{https://doi.org/10.1088/1748-0221/19/11/P11021}
		
		\bibitem{STCFL1trigger}
		W. Dong et al., J. Instrum. 17, P10027 (2022).
		\url{https://doi.org/10.1088/1748-0221/17/10/P10027}
		
		\bibitem{STCFL1trigger2}
		Y. Hao et al., IEEE Trans. Nucl. Sci. 72, 429--437 (2024).
		\url{https://doi.org/10.1109/TNS.2024.3503068}
		
		\bibitem{PAT1}
		D.E. Knuth, J.H. Morris Jr., V.R. Pratt, SIAM J. Comput. 6, 323--350 (1977).
		\url{https://doi.org/10.1137/0206024}
		
		\bibitem{PAT2}
		A.V. Aho, M.J. Corasick, Commun. ACM 18, 333--340 (1975).
		\url{https://doi.org/10.1145/360825.360855}
		
		\bibitem{PAT3}
		C.M. Hoffmann, M.J. O'Donnell, J. ACM 29, 68--95 (1982).
		\url{https://doi.org/10.1145/322290.322295}
		
		\bibitem{PAT4}
		J.R. Ullmann, J. ACM 23, 31--42 (1976).
		\url{https://doi.org/10.1145/321921.321925}
		
		\bibitem{HubaraBNN}
		I. Hubara et al., Adv. Neural Inf. Process. Syst. 29 (2016).
		
		\bibitem{OSCAR3}
		X.C. Ai et al., Nucl. Sci. Tech. 36, 242 (2025).
		\url{https://doi.org/10.1007/s41365-025-01833-x}
		
		\bibitem{STCFBackground}
		Z. Fang et al., J. Instrum. 19, P11014 (2024).
		\url{https://doi.org/10.1088/1748-0221/19/11/P11014}
		
		\bibitem{BESIIIAging}
		M.Y. Dong et al., Nucl. Instrum. Methods Phys. Res. A 1066, 169582 (2024).
		\url{https://doi.org/10.1016/j.nima.2024.169582}
		
		\bibitem{JiaSegment}
		P. Jia et al., arXiv:2605.15577 (2026).
		
		\bibitem{KiselCBM}
		I. Kisel, Nucl. Instrum. Methods Phys. Res. A 566, 85--88 (2006).
		\url{https://doi.org/10.1016/j.nima.2006.05.040}
		
		\bibitem{JuGNN}
		X. Ju et al., Eur. Phys. J. C 81, 876 (2021).
		\url{https://doi.org/10.1140/epjc/s10052-021-09675-8}
		
		\bibitem{BelleIIGNN}
		L. Reuter et al., Comput. Softw. Big Sci. 9, 6 (2025).
		\url{https://doi.org/10.1007/s41781-025-00135-6}
		
		\bibitem{YinSTE}
		P. Yin et al., arXiv:1903.05662 (2019).
		\url{https://doi.org/10.48550/arXiv.1903.05662}
		
		\bibitem{OSCAR1}
		W.H. Huang et al., J. Instrum. 18, P03004 (2023).
		\url{https://doi.org/10.1088/1748-0221/18/03/P03004}
		
		\bibitem{OSCAR2}
		X. Ai et al., Mod. Phys. Lett. A 39, 2440006 (2024).
		\url{https://doi.org/10.1142/S0217732324400066}
		
		\bibitem{SIMT}
		NVIDIA, CUDA Programming Guide, section 1.2, Programming Model.
		\url{https://docs.nvidia.com/cuda/cuda-programming-guide/01-introduction/programming-model.html}.
		Accessed 20 March 2026
		
		\bibitem{CUDA}
		NVIDIA, CUDA Platform for Accelerated Computing.
		\url{https://developer.nvidia.com/cuda}.
		Accessed 20 March 2026
		
	\end{thebibliography}
\end{document}